\documentstyle[floats,epsbox,aps,prl]{revtex}

\begin{document}

\draft

\twocolumn[\hsize\textwidth\columnwidth\hsize\csname 
@twocolumnfalse\endcsname

\title{Possible chiral phase transition in two-dimensional solid {}$^3$He}

\author{Tsutomu Momoi, Kenn Kubo and Koji Niki}
\address{Institute of Physics, University of Tsukuba, 
Tsukuba, Ibaraki 305, Japan}

\date{Recieved 19 May 1997}

\maketitle

\begin{abstract}
We study a spin system with two- and four-spin exchange interactions on 
the triangular lattice, which is a possible model for the 
nuclear magnetism of solid ${}^3$He layers. 
It is found that a novel spin structure with scalar chiral order appears 
if the four-spin interaction is dominant. 
Ground-state properties are studied using the spin-wave approximation. 
A phase transition concerning the scalar chirality 
occurs at a finite temperature, even though the dimensionality of the 
system is two and the interaction has isotropic spin symmetry. 
Critical properties of this transition are studied with 
Monte Carlo simulations in the classical limit. 
\end{abstract}
\vskip1.5pc]
\narrowtext

Solid {}$^3$He films adsorbed on the surface of graphite are purely 
two-dimensional (2D)
magnetic systems with isotropic interactions\cite{Godfrin}. 
Recently the solid phase of the adsorbed {}$^3$He layer at a low density 
has been 
attracting extensive interest, since it is a typical example of frustrated 
quantum spin systems; nuclei of {}$^3$He form a spin-1/2 quantum 
antiferromagnet on the triangular lattice. 

In a double-layer solid {}$^3$He film, only the second layer is responsible 
for its magnetism at mK region. Specific-heat measurement of the 
second layer showed a peculiar behavior\cite{Greywall89} 
and the effective exchange coupling obtained from the susceptibility was 
revealed to be antiferromagnetic\cite{Godfrin} 
at a coverage where the layer just solidifies. 
It is believed that the second layer forms a triangular lattice. 
Elser proposed that the second layer may decompose 
into two sublattices, and atoms on only one sublattice 
which form a kagom\'e net are responsible for the magnetism\cite{Elser89}. 
Recent specific-heat data, 
however, showed that all spins on the triangular lattice contribute to the 
magnetism and they cannot be fully separated to the kagom\'e lattice 
and the other\cite{SiqueiraNCSa,Fukuyama96}. 
It is not yet clear whether the inequivalence of the two sublattices gives 
relevant effects on the translational symmetry of the magnetic interactions. 
Another example of 2D solid {}$^3$He is 
realized in a monolayer of solid {}$^3$He 
adsorbed on graphite preplated with {}$^4$He\cite{LusherSC} or 
HD\cite{SiqueiraLCS93}. 

It is well established that the multi-spin exchange interactions are 
important in the nuclear magnetism of solid {}$^3$He\cite{RogerHD}. 
These multi-spin interactions originate from particle ring-exchange 
of {}$^3$He atoms. It was shown by Thouless that 
the whole Hamiltonian has the form 
$H = \sum_n (-1)^n \sum J_n P_n$, 
where $P_n$ denotes cyclic permutation of $n$ spins and the 
exchange constant $J_n$ is always negative\cite{Thouless65}. 
According to numerical calculations for the triangular 
lattice\cite{RogerWKB,BernuCL92}, 
two- and three-spin exchanges are large and, 
furthermore, four- and six-spin exchanges are not small. 
As pointed out in Ref.~\onlinecite{Roger90}, 
the multi-spin interactions can create frustration, and hence 
they should be properly considered in a theoretical treatment. 
In our previous study\cite{KuboM}, we found 
that the four-spin interaction can produce novel phases with 
four-sublattice structures. 

In this Letter, we reveal effects of the four-spin interaction 
and predict a chiral phase transition in a certain 
parameter region where four-spin interactions are dominant. 
It is found that the ground state of a spin model for the solid ${}^3$He layer 
has a scalar chiral order. The spin-wave approximation 
shows that quantum fluctuations are not strong enough to destroy 
the chiral order at $T=0$. We hence expect that the real 
{}$^3$He layer shows a chiral phase transition at a finite temperature. 
Critical properties of this transition are also studied 
in the classical limit. The critical exponent $\alpha$ of the 
specific heat clearly deviates 
from that of the 2D Ising model. 
To our knowledge, the present model 
gives the first example of a chiral phase transition in a 2D 
realistic spin system with $SO(3)$ symmetry. 

We consider a spin model with two-, three- and four-spin exchange 
interactions on the triangular lattice, 
and assume that the interactions have the same 
translational symmetry as that of the triangular lattice. 
Since the three-spin permutations can be transformed into 
two-spin exchanges, the Hamiltonian can be written as 
\begin{equation}\label{Hamiltonian}
H= J \sum_{\langle i,j \rangle} 
        \mbox{\boldmath $\sigma$}_i \cdot \mbox{\boldmath $\sigma$}_j 
  + K \sum_p h_p,
\end{equation}
where $J=J_3-J_2/2$, $K=-J_4/4(\ge 0)$, and $\mbox{\boldmath $\sigma$}_i$ 
denote Pauli matrices. 
The first and the second summations run over all pairs of nearest neighbors 
and all minimum diamond clusters, respectively. 
The explicit form of $h_p$ for four sites $(1,2,3,4)$ is 
\begin{eqnarray}
 h_p &=& \sum_{1\le i<j \le 4} 
            \mbox{\boldmath $\sigma$}_i \cdot \mbox{\boldmath $\sigma$}_j 
      + (\mbox{\boldmath $\sigma$}_1 \cdot \mbox{\boldmath $\sigma$}_2)
        (\mbox{\boldmath $\sigma$}_3 \cdot \mbox{\boldmath $\sigma$}_4) 
\nonumber \\
     &+& (\mbox{\boldmath $\sigma$}_1 \cdot \mbox{\boldmath $\sigma$}_4)
         (\mbox{\boldmath $\sigma$}_2 \cdot \mbox{\boldmath $\sigma$}_3) 
      - (\mbox{\boldmath $\sigma$}_1 \cdot \mbox{\boldmath $\sigma$}_3)
        (\mbox{\boldmath $\sigma$}_2 \cdot \mbox{\boldmath $\sigma$}_4),
\end{eqnarray}
where $(1,3)$ and $(2,4)$ are diagonal bonds of the diamond. 
Bernu et al.\ first studied this model discussing the temperature 
dependence of 
the specific heat and the magnetic susceptibility\cite{BernuCL92}. 
Roger estimated density dependence 
of interactions using the WKB approximation\cite{RogerWKB}. 
It shows that $J_3$ is dominant in the high-density region and 
$|J_2|$ increases faster than $|J_3|$ 
as the density is lowered. Hence the value 
of $J$ changes rapidly depending on the density 
and $J$ may vanish at a low-density region. 

In the previous paper\cite{KuboM}, we studied the ground state of the 
Hamiltonian (\ref{Hamiltonian}) for various $J$ with 
the mean-field approximation. For $J<-8K$, the ground state shows 
the perfect ferromagnetism. 
For $-8K<J<-8K/3$, ground states have non-trivial degeneracy. 
One of the ground states has a four-sublattice 
structure, in which three spins are up and the other is down. 
There exist other kinds of ground states 
with longer periods. In $-8K/3<J<25K/3$, the ground state has 
a four-sublattice structure with zero magnetization, which we call as 
the tetrahedral structure (see Fig.~1). This spin configuration 
is indeed proved to be the exact ground state for the finite region 
$-K/2\le J \le 2K$ in the classical limit. (See Ref. \cite{KuboM}.) 
For $25K/3 <J$, the ground state is 
the so-called $120^\circ$ structure. Thus novel phases appear due to 
the four-spin interactions. Among these states, the tetrahedral structure 
has an interesting property; it has the scalar chiral order 
and hence the ground-state manifold has two-fold degeneracy. 
In this Letter, we employ the following order-parameter operator 
of the chirality\cite{WenWZ} 
\begin{equation}
\kappa = \sum_{(i,j,k)} \mbox{\boldmath $\sigma$}_i \cdot 
           (\mbox{\boldmath $\sigma$}_j \times \mbox{\boldmath $\sigma$}_k),
\end{equation}
where the summation runs over all unit triangles 
and the indices $(i,j,k)$ are chosen in clockwise turn for each triangle. 
\begin{figure}
\begin{center}
  \epsfile{file=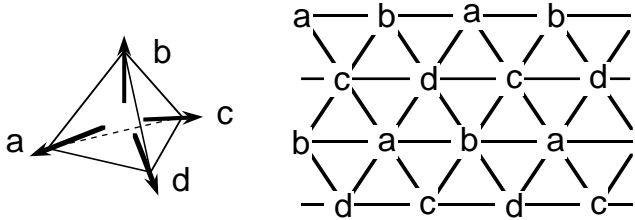,width=3.36in}
\end{center}
\caption{Spin vectors of four sublattices in the tetrahedral
structure (left) and the configuration of the sublattices $a,b,c,d$ on the
triangular lattice (right).}
\label{fig1}
\end{figure}

To study the chiral order, we concentrate on the case only with 
the four-spin interaction, i.e., $J=0$. 
This system has strong frustration due to the four-spin exchange and 
hence quantum effects should be properly treated. 
We employ the spin-wave approximation to investigate the strength 
of quantum fluctuations. Using the Holstein-Primakoff transformation, we 
expand the Hamiltonian in terms of Bose operators up to bilinear terms 
from the tetrahedral structure. 
After the Bogoliubov transformation of four kinds of bosons, we obtain 
\begin{equation}
H = - {49 \over 3}KN 
    + \sum_k \sum_{i=1}^4 \omega_i (\mbox{\boldmath $k$}) 
(a_{ik}^\dagger a_{ik} + {1 \over 2}), 
\end{equation}
where the first summation runs over the reduced Brillouin zone of the 
four-sublattice structure. The frequency $\omega_1 (\mbox{\boldmath $k$})$ 
is given by 
\begin{equation}
\omega_1 (\mbox{\boldmath $k$}) 
   = {8K\over 9} (4{A_k}^2 - {B_k}^2 - 3{C_k}^2)^{1/2} ,
\end{equation}
where
\begin{eqnarray}
A_k &=& 12 - \cos {k_1 \over 2} 
        - \cos \bigl({k_1 \over 4} + {\sqrt3 k_2 \over 4}\bigr) \nonumber\\
       &-& \cos \bigl({k_1 \over 4} - {\sqrt3 k_2 \over 4}\bigr) 
        + \cos \bigl({3 k_1 \over 4} + {\sqrt3 k_2 \over 4}\bigr) \nonumber\\
       &+& \cos {\sqrt3 k_2 \over 2}  
        + \cos \bigl({3 k_1 \over 4} - {\sqrt3 k_2 \over 4}\bigr), \\
B_k &=& 5 \Bigl\{ - \cos {k_1 \over 2} 
        + 2 \cos \bigl({k_1 \over 4} + {\sqrt3 k_2 \over 4}\bigr) \nonumber\\
       &-& \cos \bigl({k_1 \over 4} - {\sqrt3 k_2 \over 4}\bigr) \Bigr\}
        - \cos \bigl({3 k_1 \over 4} + {\sqrt3 k_2 \over 4}\bigr) \nonumber\\
       &-& \cos {\sqrt3 k_2 \over 2} 
        + 2 \cos \bigl({3 k_1 \over 4} - {\sqrt3 k_2 \over 4}\bigr), \\
C_k &=& 5 \Bigl\{ \cos {k_1 \over 2} 
        - \cos \bigl({k_1 \over 4} - {\sqrt3 k_2 \over 4}\bigr) \Bigr\} 
\nonumber\\
    &-& \cos \bigl({3 k_1 \over 4} + {\sqrt3 k_2 \over 4}\bigr) 
     + \cos {\sqrt3 k_2 \over 2}.
\end{eqnarray}
The others are $\omega_2(\mbox{\boldmath $k$})
=\omega_1(\mbox{\boldmath $k$}+\mbox{\boldmath $q$}_1)$, 
$\omega_3(\mbox{\boldmath $k$})
=\omega_1(\mbox{\boldmath $k$}+\mbox{\boldmath $q$}_2)$ and 
$\omega_4(\mbox{\boldmath $k$})=
\omega_1(\mbox{\boldmath $k$}+\mbox{\boldmath $q$}_1+\mbox{\boldmath $q$}_2)$, 
where $\mbox{\boldmath $q$}_1$ and $\mbox{\boldmath $q$}_2$ are reciprocal 
lattice vectors of the four-sublattice structure.  
Three frequencies $\omega_i (\mbox{\boldmath $k$})$ with $i=2,3,4$ 
are gapless at $\mbox{\boldmath $k$}=0$ 
and the other one, $\omega_1 (\mbox{\boldmath $k$})$, is massive. 
The ground-state energy per site is estimated as 
$\varepsilon_{\rm g} = -7.4615K$, whereas $\varepsilon_{\rm g}=-17K/3$ 
in the classical limit. 
The sublattice magnetization 
and the chirality are evaluated in the same way.  
The estimate of the sublattice magnetization per spin is $m_{\rm s}/S=0.5937$, 
where $S=1/2$. 
The deviation from the classical value, $\Delta m_{\rm s}/S=0.4063$, 
is small compared with the value $\Delta m_{\rm s}/S=0.522$ of 
the pure Heisenberg antiferromagnet on the triangular 
lattice\cite{ThJolicoeurG}, which 
indicates that quantum fluctuations are not very strong and the ground 
state has the sublattice magnetic order. The value of the chirality per 
unit triangle is $\kappa/2N=0.2858$, 
whereas $\kappa/2N=4/3\sqrt3$ in the classical limit. 
The chirality has 37\% of its classical value and this ratio is larger than 
the third power of the sublattice magnetization. 
This shows that the chiral order is more stable against 
quantum fluctuations than the sublattice magnetization. 
A similar tendency was 
observed in the vector chiral order of the Heisenberg and $XY$ 
antiferromagnets on the triangular lattice\cite{MomoiS}. 
The spin-wave analysis of the {\it uuud} and $120^\circ$ structure ground
states indicates 
that the stable region for the tetrahedral ground state will become narrower 
than the mean field result\cite{KuboNM}. 

\begin{figure}
\begin{center}
  \epsfile{file=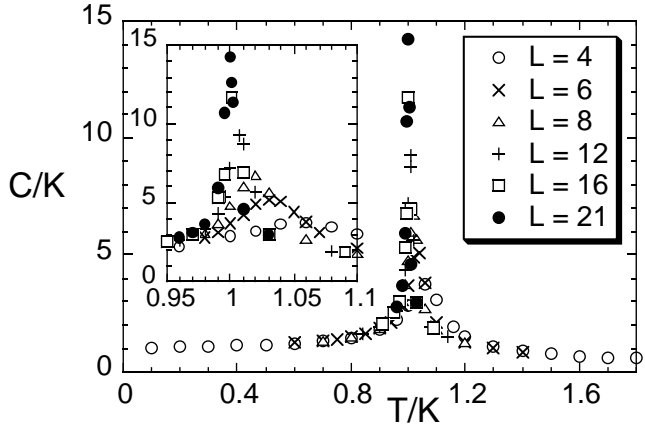,width=3.36in}
\end{center}
\caption{Temperature dependence of the specific heat at $J=0$. 
The figure around the peak is extended in the inset. }
\label{fig2}
\end{figure}
The greatest significance of the chiral order is that it 
can exist even at a finite temperature since it is stable 
against the spin-wave fluctuations with long wave-length. 
The spin-wave 
approximation indeed shows that the estimate of the chiral order is 
nonvanishing at sufficiently low temperatures. 
We thus expect a phase transition 
to occur at a finite temperature, which is accompanied by 
ordering of the scalar chirality. We study the finite-temperature 
properties in the classical limit using Monte Carlo simulations. 
We believe that critical properties of the phase transition 
at finite temperatures are governed by thermal fluctuations and 
hence the quantum effects do not change its universality class, 
though they can change values of $T_{\rm c}$ and the order parameter. 
We treat $\mbox{\boldmath $\sigma$}_i$ as a classical unit vector. 
Monte Carlo simulations are performed with 
the Metropolis algorithm. If a spin flip is rejected, we randomly 
rotate the spin about the local molecular field. 
We construct finite-size systems with a unit cluster which has 12 sites. 
The system size is $N=12 L^2$ with $L=4$, 6, 8, 12, 
16, 21 with a periodic-boundary condition. 
After discarding initial 6000--50000 Monte Carlo steps per spin (MCS) 
for equilibration, subsequent $12\times 10^4$--$5\times 10^5$ MCS are 
used to calculate the 
average. Further details will be reported in Ref.~\cite{MomoiK}. 

\begin{figure}
\begin{center}
  \epsfile{file=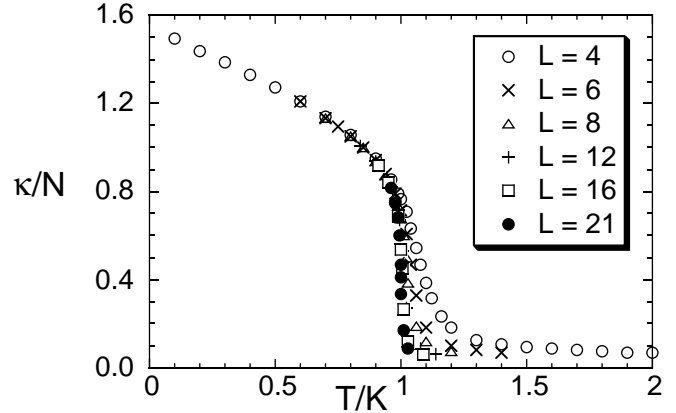,height=5.6cm}
\end{center}
\caption{Temperature dependence of the scalar chirality at $J=0$. }
\label{fig3}
\end{figure}
The specific-heat data (Fig.~2) show a sharp peak around 
$T_{\rm p}=1.00K \sim 1.05K$ and this 
peak diverges as $L$ increases. 
We verified that this phase transition is not of first order, by 
studying the cumulant of the fourth moment of the energy 
$1-\langle E^4 \rangle / 3\langle E^2 \rangle^2$. 
The response function of the chirality, $\chi = \langle \kappa^2 \rangle /NT$, 
shows strong divergence at $T_{\rm p}$ and the long-range order 
of the chirality, 
$\sqrt{ \langle \kappa^2 \rangle } /N$, 
appears below this temperature (Fig.~3). 
Of course, there is no sublattice magnetic order at finite temperatures 
even in the chiral ordered phase. 
Thus the divergence of the 
specific heat corresponds to the second-order phase transition 
from the disordered phase
to the chiral ordered one. This is the first example of the scalar-chiral 
phase transition in the 2D realistic spin model with $SO(3)$ 
symmetry. We estimate the critical exponent $\alpha$ of the 
specific heat, $C(T)\sim |T-T_{\rm C}|^{-\alpha}$. 
The maximum values of the specific heat are plotted in Fig.~4 for 
various sizes. The finite-size scaling 
reveals that the peak value behaves as $C(T_{\rm p}(L)) \sim L^{\alpha/\nu}$. 
The plot fits well by setting $\alpha/\nu=0.9(1)$. 
We hence obtain $\alpha=0.62(5)$ and $\nu=0.69(3)$ 
using the scaling relation 
$\alpha+2\nu=d$. This is clearly different from the $log$-divergence, 
i.e. $\alpha=0$, of the 2D Ising model. 
This result might be thought peculiar, 
since the chirality is an Ising-type variable and 
the chiral phase transition is expected to have the Ising 
universality. 
There is however a similar example; the four-spin interaction added 
to the Ising model changes the exponent $\alpha$ continuously from 
0\cite{Ising}. 
The four-spin interaction may have a universal effect to change the 
critical exponent. 
Other values are estimated as $\beta/\nu=0.13(2)$, $\gamma/\nu=1.75(5)$ and 
$T_{\rm c}=0.9935(10)$ from the finite-size scaling 
of the order parameter and the response function\cite{MomoiK}. 
The critical indices may depend on the parameter $J/K$ as
they do in the 2D Ising model with four-spin interactions. 
Our estimates of the exponents 
$(2-\alpha)/\nu$, 
$\beta/\nu$, and $\gamma/\nu$ appear to equal to 2, 1/8, and 7/4, 
respectively, and hence the present transition still belongs to the Ising 
universality-class in the sense of the weak universality\cite{Suzuki}. 
\begin{figure}
\begin{center}
  \epsfile{file=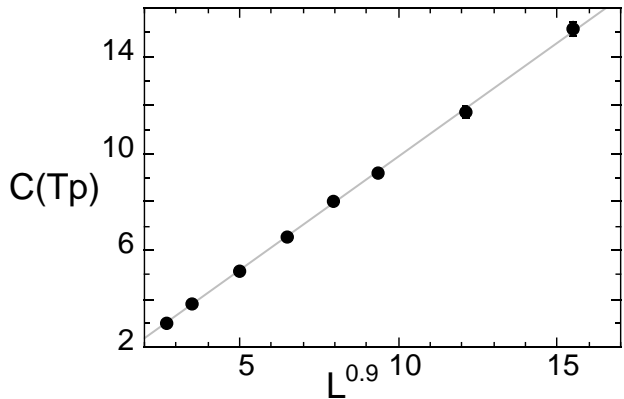,width=3.25in}
\end{center}
\caption{Finite-size scaling plot of the peak values of the specific heat.} 
\label{fig4}
\end{figure}

Even for the vector chirality in the fully frustrated 
2D $XY$ model, there remain controversies whether the chiral 
phase transition has the Ising universality or not. Since Miyashita and 
Shiba\cite{MiyashitaS} estimated $\alpha=0$ using Monte Carlo simulations, 
this problem has been studied repeatedly\cite{v_chiral}. 
Though the discussions have not yet settled down, 
all estimates of $\alpha$ are less than $0.4$. 
Our estimate is clearly different from these values. 

We mention the possibility that the chiral phase transition may be 
observed experimentally in the solid ${}^3$He films. 
The ${}^3$He films behave as ferromagnets in a high-density 
region\cite{Godfrin,Schiffer93,Morishita96} and, by lowering the
coverage, the exchange coupling $J_\chi$ derived from the 
susceptibility turns antiferromagnetic\cite{Godfrin,SiqueiraNCSb}. 
On the other hand, the high-temperature series expansion shows that 
$J_\chi = -2(J+6K)$\cite{BernuCL92}.
Hence the negativeness of $J_\chi$ indicates that 
$J>-6K$ and not necessary that $J$ is antiferromagnetic. 
For all densities, no magnetic phase transition has been observed 
at finite temperatures\cite{Morishita96,SiqueiraNCSb}. 
Comparing these results with the phase diagram of the present spin 
model\cite{KuboM}, we find that the low-density ${}^3$He layer seems 
to correspond to the 
intermediate phase ($-8K<J<-8K/3$) between the ferromagnetic phase 
and the tetrahedral-structure phase. 
According to the WKB approximation\cite{RogerWKB}, the value of 
$|J|(=|J_2/2 - J_3|)$ decays fast whereas $K$ increases as the 
density is lowered. We hence expect that, in a certain lower-density 
(but well solidified) region, the four-spin exchange 
interaction is dominant and the chiral phase 
transition may occur. 
At this phase transition, one can observe sharp divergence 
of the specific heat, as we have shown in this Letter, 
and a cusp-like behavior in the magnetic susceptibility 
(see Ref.~\onlinecite{MomoiK}). 
Recently, it was reported that, by preplating HD on graphite, the solid 
phase of a monolayer of {}$^3$He can be stabilized and then the {}$^3$He film 
can first solidify around 
$\rho=0.0555 \AA^{-2}$\cite{SiqueiraLCS93,SiqueiraNCSb}, 
which is the lowest density ever observed. 
(The second layer of double-layer {}$^3$He solidifies 
around $\rho_2=0.064 \AA^{-2}$.) 
We expect that this material may be a plausible 
candidate for showing the chiral phase transition. 

The authors wish to thank Hiroshi Fukuyama, Hikaru Kawamura and 
Harumi Sakamoto for stimulating discussions and useful comments. 
This work was supported by Grant-in-Aid No.\ 09440130 from Monbu-shou and 
T.~M. was supported in part by the Japan Society for the Promotion of 
Science (JSPS). 
The numerical calculations were done on Facom VPP500 at the ISSP of the 
University of Tokyo.


\end{document}